\documentclass{jaa}
\usepackage{txfonts}
\usepackage{graphicx}
\newcommand{\gtsim}{\stackrel >{_\sim}}
\newcommand{\ltsim}{\stackrel <{_\sim}}

\begin{document}

\title[GMRT Galactic HI Absorption survey II]
{A High Galactic Latitude HI 21cm-line Absorption Survey using the GMRT:
                         II. Results and Interpretation }
\author[Mohan, Dwarakanath \& Srinivasan]
       {Rekhesh Mohan\thanks{Currently at the Indian Institute of Astrophysics,
                             Bangalore 560 034, India.}
       \thanks{e-mail:reks@iiap.res.in},
       K. S. Dwarakanath\thanks{e-mail:dwaraka@rri.res.in},
       \& G. Srinivasan\thanks{e-mail:srini@rri.res.in} \\
        Raman Research Institute, Bangalore 560 080, India }
\maketitle
\label{firstpage}

\begin{abstract}
We have carried out a sensitive high-latitude ($|b|$ $>$ 15$^{\circ}$) HI
21cm-line absorption survey towards 102 sources using the GMRT. With a
$3\sigma$ detection limit in optical depth of $\sim$0.01, this is the most 
sensitive HI
absorption survey. We detected 126 absorption features most of which also
have corresponding HI emission features in the Leiden Dwingeloo Survey of
Galactic neutral Hydrogen.
The histogram of random velocities of the absorption features is
well-fit by two Gaussians centered at V$_{lsr}$ $\sim$ 0 km s$^{-1}$ with
velocity dispersions of 7.6 $\pm$ 0.3 km s$^{-1}$ and 21 $\pm$ 4 km s$^{-1}$
respectively. About 20\% of the HI absorption features form the larger
velocity dispersion component.
The HI absorption features forming the narrow Gaussian have a mean
optical depth of 0.20 $\pm$ 0.19, a mean HI column density of
(1.46 $\pm$ 1.03) $\times$ 10$^{20}$ cm$^{-2}$, and a mean
spin temperature of 121 $\pm$ 69 K. 
These HI concentrations can be identified
with the standard HI clouds in the cold neutral medium of the Galaxy.
The HI absorption features forming the wider Gaussian have a mean optical
depth of 0.04 $\pm$ 0.02, a mean HI column density of
(4.3 $\pm$ 3.4) $\times$ 10$^{19}$ cm$^{-2}$, and a mean spin
temperature of 125 $\pm$ 82 K. 
The HI column densities of these fast clouds
decrease with their increasing random velocities. These fast clouds can be
identified with a population of clouds detected so far only in optical 
absorption and in HI emission lines with a similar velocity dispersion. This
population of fast clouds is likely to be in the lower Galactic Halo.
\end{abstract}


\section{Introduction}
\label{sec:intro}
A number of HI 21cm-line absorption surveys have been carried out in the
last 50 years or so. While interferometric surveys are a better
alternative for such studies, since it rejects the more extended HI emission,
the lack of collecting area of the interferometers
has limited the sensitivity of the HI absorption
surveys. Until recently, the Very Large Array (VLA) used to be the only
instrument with a collecting area comparable to large single
dish telescopes. From the various HI absorption surveys carried out so
far, more than 600 absorption spectra are available, but the optical depth
detection limits of more than 75\% of these are above 0.1 (see for eg.
Mohan et al, \nocite{me2004}2004 - hereafter paper I, for a summary of
previous surveys).

The motivation for the present survey, the observing strategy, the sources
observed, their HI 21cm line spectra and the parameters of the discrete HI line
components are presented in \nocite{me2004}paper I. Here, we discuss their 
interpretation.
As was mentioned in \nocite{me2004}paper I, the low optical depth regime 
of Galactic HI is
largely unexplored, except for the HI absorption study by
Dickey et al \nocite{dst78}(1978) and by Heiles \& Troland (\nocite{ht03a}2003a,
\nocite{ht03b}b). Dickey et al. \nocite{dst78}(1978) measured HI 
absorption/emission towards 27 extragalactic
radio sources located at high and intermediate Galactic latitudes
($|b|$ $>$ 5$^{\circ}$) using the Arecibo telescope. The rms optical depth
in their spectra was $\sim$0.005.
These profiles can be considered the best in terms of signal-to-noise ratio,
though in many of the profiles the systematics in the band dominate
the noise. Despite these limitations, Dickey et al
\nocite{dst78}(1978) noted that the velocity distribution of HI absorption
features is dependent on their optical depths. For the optically thin clouds
($\tau$ $<$ 0.1), the velocity dispersion was $\sim$11 km s$^{-1}$,
whereas for the optically thick clouds ($\tau$ $>$ 0.1) this value was
$\sim$6 km s$^{-1}$. Similar trend was also noticed in the
Effelsberg-Green Bank survey (Mebold et al, \nocite{egs82}1982). More recently,
Heiles \& Troland \nocite{ht03b}(2003b) also found indications for an
independent population of lower optical depth HI absorption features. 

\begin{figure}
\begin{center}
\includegraphics[width=7.5cm,height=13.0cm,angle=-90]{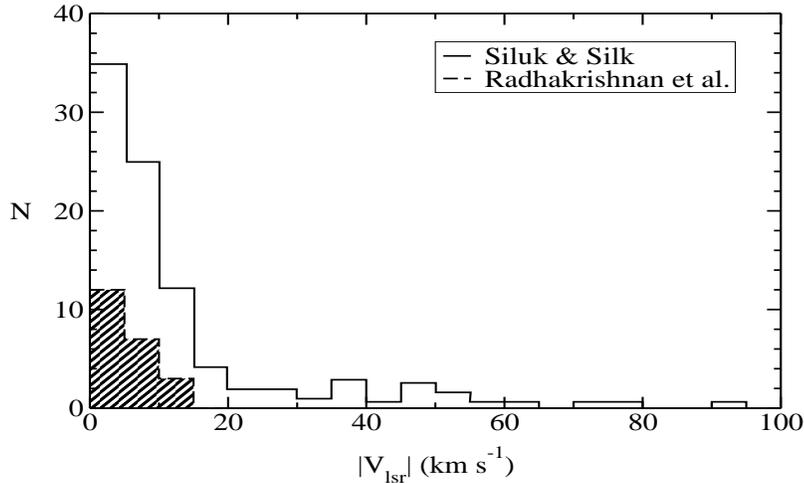}
\end{center}
\caption{Histogram of radial (random) velocities from the optical
absorption line studies by Siluk \& Silk (1974), along with that
from the
HI absorption line survey by Radhakrishnan et al., 1972
(dashed line; shaded). The higher velocity tail is missing in the radio
observations.}
\label{fig:ss74rad72}
\end{figure}

One of the first efforts to study the nature of the ISM was the observation
of interstellar absorption in the optical line of singly ionized calcium
(CaII) towards early type stars by Adams \nocite{adams}(1949). 
He noted that the observed extent in the radial velocities exceeded 50
km s$^{-1}$ in the local standard of rest frame (LSR). Blaauw \nocite{b52}(1952)
analysed the random velocity distribution of interstellar absorption lines
in Adams' data. One of the main conclusions of this analysis was that the 
random velocity distribution of
interstellar absorption features cannot be explained by a single Gaussian
distribution. Support for this conclusion came from the study of Routly \& 
Spitzer \nocite{rs52}(1952), who found the ratios of column densities
of neutral sodium (NaI) to singly ionized calcium (CaI) to decrease 
systematically with increasing random
velocities of the absorption features. This effect, called the ``Routly-Spitzer
effect'' was later confirmed from a much larger sample of stars by 
Siluk \& Silk \nocite{ss74}(1974).
Field, Goldsmith \& Habbing  \nocite{fgh69}(1969) modeled the ISM as
cool dense concentrations of gas, often referred to as 
``interstellar clouds'' (the Cold Neutral Medium or CNM), in pressure 
equilibrium with a warmer intercloud medium (the Warm Neutral Medium or WNM). 
While this initial model of the ISM has been refined considerably by
later studies, the basic picture of the ISM with cold diffuse clouds
and the warmer intercloud medium has remained.
However, there has been a discrepancy in the velocity distribution of 
interstellar clouds (the CNM) obtained from the optical absorption line studies
and from the HI 21-cm line observations. The optical as well as UV
absorption line studies indicated the presence of features with larger
spread in random velocities, which was absent in the 21-cm line observations 
(Mohan et al, \nocite{me2001}2001 \& the references therein). This is clear 
from Fig. \ref{fig:ss74rad72}.

According to one hypothesis, the high
velocity ($|v|$ $\gtsim$ 15 km s$^{-1}$) optical absorption lines arise
in interstellar clouds, shocked and
accelerated by supernova remnants in their late phases of evolution
(Siluk \& Silk, \nocite{ss74}1974; Radhakrishnan \& Srinivasan,
\nocite{rs80}1980). Such a
mechanism would naturally result in the higher random velocity clouds being
warmer and having lower column densities due to shock heating and
evaporation, as compared to the low random
velocity clouds. Lower column density
explains the non detections of such clouds in HI emission, since the intensity
of the HI 21cm line emission is directly proportional
to the HI column density. Higher temperature, along with lower column
density ($N_{\scriptscriptstyle{\rm{HI}}}$) results in lower optical depth, 
since $\tau_{\scriptscriptstyle{\rm{HI}}}$ $\propto$
$N_{{\scriptscriptstyle{\rm{HI}}}}/T$,
where $T$ is the excitation temperature of the
spectral line (Siluk \& Silk, \nocite{ss74}1974;
Radhakrishnan \& Srinivasan, \nocite{rs80}1980;
Rajagopal et al, \nocite{jd98b}1998). If this scenario is correct, then a
sensitive HI absorption survey should detect those features with lower
optical depth and higher random velocities which are presumably the
counterparts of the higher velocity optical absorption lines. We have used the
present dataset to address this scenario.

In the next section we discuss a method to estimate the
contributions from the Galactic differential rotation to the observed radial
velocities of the HI absorption features.
To supplement the absorption spectra, the corresponding HI 21cm emission
profiles from the Leiden-Dwingeloo HI emission survey (LDS, Hartman \& Burton,
\nocite{lds}1995) were analyzed. Appendices B \& C of paper I list the HI line 
profiles (absorption \& emission), and their fitted line parameters respectively.
In section \ref{sec:old_radio+opt}, we compare the present dataset with the
previous HI absorption surveys. 
In Section \ref{sec:abs}, we discuss the statistics of HI absorption
line parameters obtained from the GMRT. The
velocity dispersion of the interstellar clouds estimated from the present
survey and the low optical depth features are
highlighted. In section \ref{sec:discuss}, we compare our results with various
other existing studies to discuss
the location of the newly detected high random velocity and low optical depth
features. 

\section{The differential Galactic rotation}
\label{sec:galrot}
As was discussed in paper I, we carried out a survey at higher Galactic latitudes
to avoid the blending of components in the absorption spectra and to minimize 
the contribution of the systematic velocities in the observed radial velocities 
due to the Galaxy's differential rotation. 
For a given Galactic longitude ($l$) and latitude ($b$), the
observed radial component of differential Galactic rotation for objects
in the solar neighbourhood is (Burton, \nocite{wbb88}1988)

\begin{equation}
{\rm{v}}_{\scriptscriptstyle{\rm{r}}} = Ar~sin~2l~~cos~b
\label{eqn:nearbygrot}
\end{equation}

where, $A$ = 14 km s$^{-1}$ Kpc$^{-1}$ is the Oort's constant and
$r$ is the heliocentric distance to the object.

\begin{figure}
\begin{center}
\includegraphics[width=7.0cm,height=10.0cm,angle=-90]{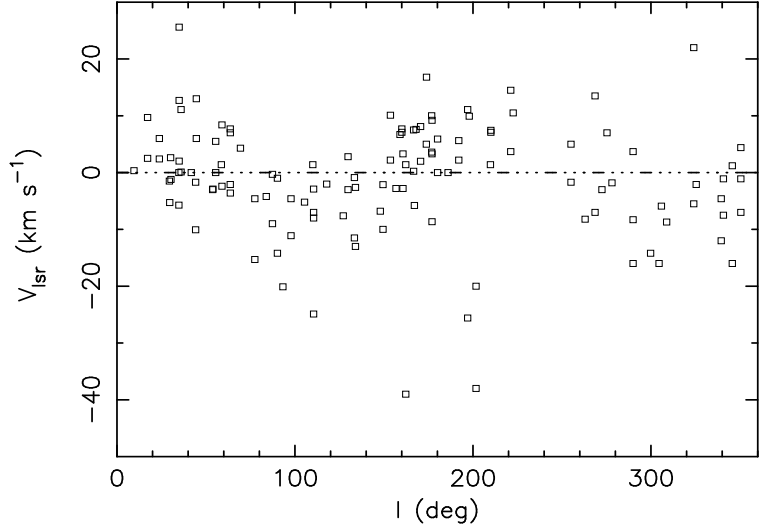}
\end{center}
\caption{The radial velocities (in the LSR frame) of the HI absorption features
detected  in the present survey as a function of Galactic longitude}
\label{fig:lvplot}
\end{figure}

In figure \ref{fig:lvplot} we have plotted the radial velocities with respect
to the local standard of rest (LSR) of the various HI absorption features as
a function of the
Galactic longitude. If the contribution to radial velocities from the
Galactic differential rotation is dominant, there should be a pronounced
signature of a ``sine wave'' in such a plot, provided the absorption features
are at the same heliocentric  distance $r$ (Eqn. \ref{eqn:nearbygrot}).
There is a suggestion that absorption features with $|V_{lsr}|$ $\ltsim$
15 km s$^{-1}$ may have significant contribution from differential rotation.

Including an additional term for the random motion of the cloud, equation
\ref{eqn:nearbygrot} becomes,

\begin{equation}
{\rm{v}}_{\scriptscriptstyle{\rm{r}}} = {\rm{v}}
_{\scriptscriptstyle{\rm{random}}} ~+~ Ar~sin~2l~~cos~b
\\
\end{equation}

It would appear that this function can be fitted to the observed distribution
(Fig. \ref{fig:lvplot}) to estimate the systematic component. However, the
distances to the absorbing features are unknown. Consider an ensemble of
interstellar clouds, each with its own random motion and distance. The
random motions can be approximated by a Gaussian with a
zero mean. If we assume the number density of clouds
to be only a function of height, $|z|$, from the Galactic plane and the Sun to
be in the plane, the distances for a distribution of clouds in the Galaxy can
be assumed to be an exponential deviate with a scale height (Dickey \& Lockman,
\nocite{araa90}1990). Adding these two terms, the above equation can be
re-written as

\begin{equation}
{\rm{v}}_{\scriptscriptstyle{\rm{r}}} = a_{\scriptscriptstyle{\rm{1}}}
{\rm{v}}_{\scriptscriptstyle{\rm{gauss}}}
~+~ a_{\scriptscriptstyle{\rm{2}}}\left(\frac{z_{\scriptscriptstyle
{\rm{exp}}}} {|sin~b|}\right)~sin~2l~~cos~b \\
\end{equation}

Where $z_{\scriptscriptstyle{\rm{exp}}}$ is an Exponential deviate
with unit scale height and ${\rm{v}}_{\scriptscriptstyle{\rm{gauss}}}$
is a Gaussian deviate. The term
$z_{\scriptscriptstyle{\rm{exp}}}/|sin~b|$ is to take into account the
variation of path length through the disk as a function of the Galactic
latitude. We have carried out a Monte-Carlo simulation to generate
such a distribution and compare with the observed distribution (Fig.
\ref{fig:lvplot}). The simulation used the Monte Carlo routines discussed by
Press et al \nocite{nrp}(1992).

We used the $F-test$ (Press et al, \nocite{nrp}1992) to check if the
simulated and the observed data can be derived from the
same distribution. We found no well defined peak
for the probability distribution. The 3$\sigma$ level indicates
that similar confidence levels can be achieved even with $a_1$ or $a_2$ = 0.0.
Therefore, the $F-test$ results imply that a systematic pattern in the
distribution is negligible (For a detailed discussion see Mohan, 
\nocite{me2003}2003). Hence, we have not applied
any correction to the observed velocity for differential Galactic rotation.
The observed radial velocities are considered to be random motions.
A similar conclusion was also reached by Heiles \& Troland 
(\nocite{ht03b}2003b) for
the recent Arecibo survey of HI absorption/emission measurements for Galactic
latitudes $|b|$ $>$ 10$^{\circ}$.

\section{Comparison of the GMRT data with the previous HI and optical surveys}
\label{sec:old_radio+opt}
In this section we
compare the histogram of random velocity distribution obtained from the
present survey with those from the previous HI absorption line surveys and 
optical absorption line surveys of interstellar NaI and CaII. For the case of
radio surveys, we have used those lines of sight from the previous surveys 
with $|b|$ $>$ 15$^{\circ}$. While comparing our data with the results 
obtained from optical surveys, we have not put this restriction since
the stars observed are located in the solar neighbourhood and the contribution from 
the differential Galactic rotation (if any) would be negligible. We have,
however, carefully analysed the optical surveys to exclude those lines
of sight where the velocities of the absorption lines were due to
systematic motions.

\subsection{Comparison of the GMRT data with the previous HI surveys}
\label{sec:highb_oldies}
\begin{figure}
\begin{center}
\includegraphics[scale=0.212,angle=-90]{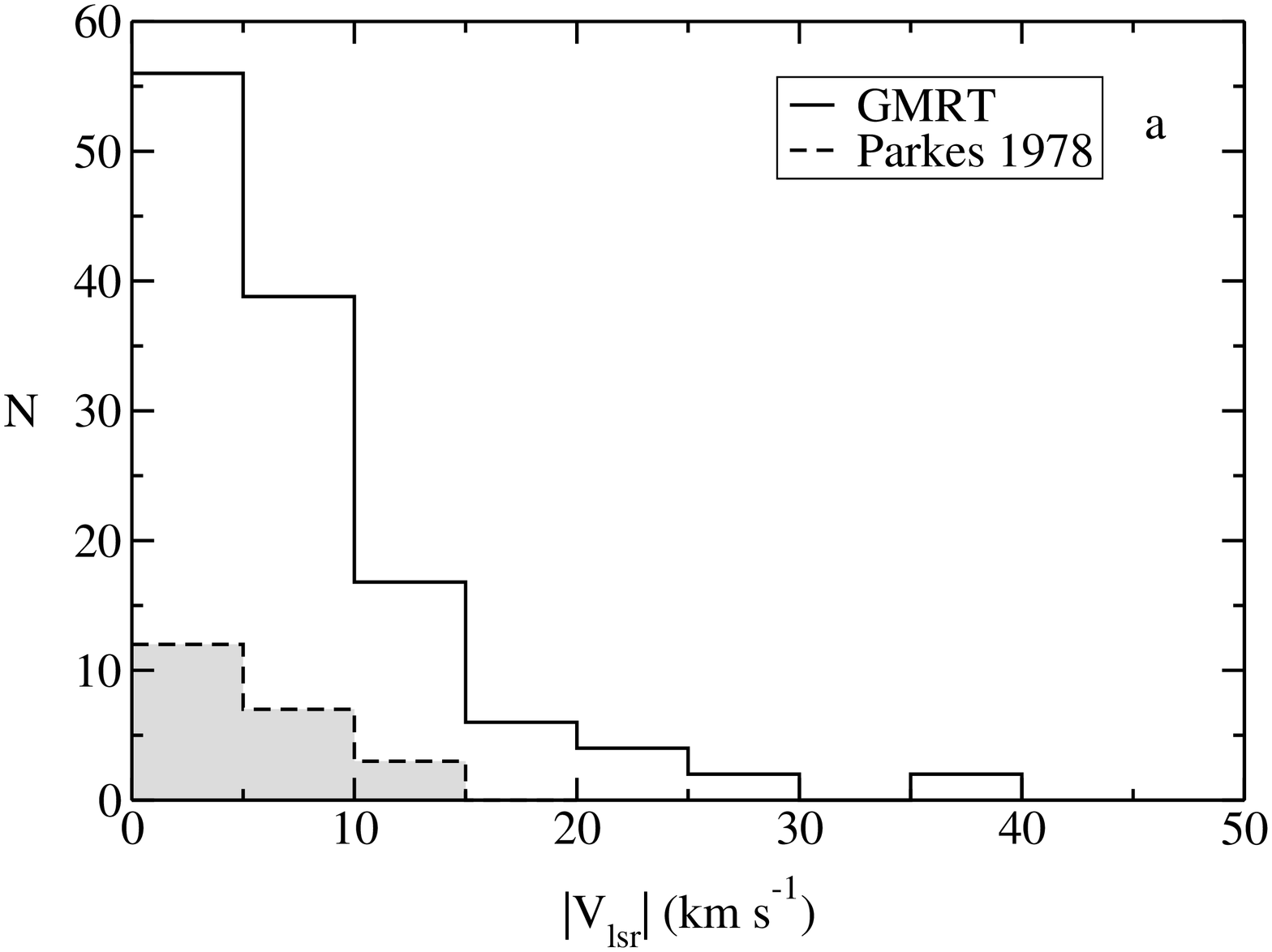}
\includegraphics[scale=0.212,angle=-90]{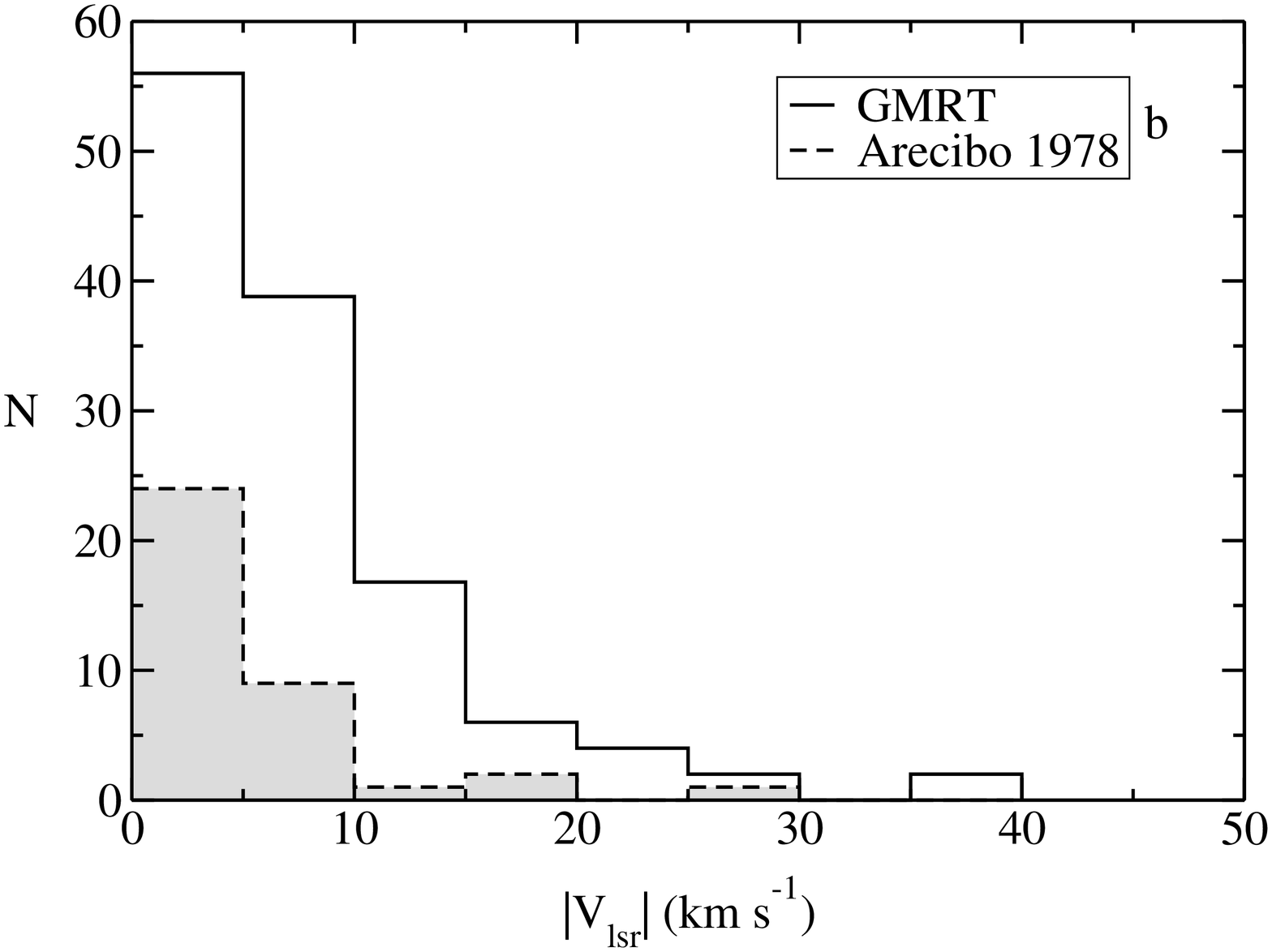}\\
\includegraphics[scale=0.212,angle=-90]{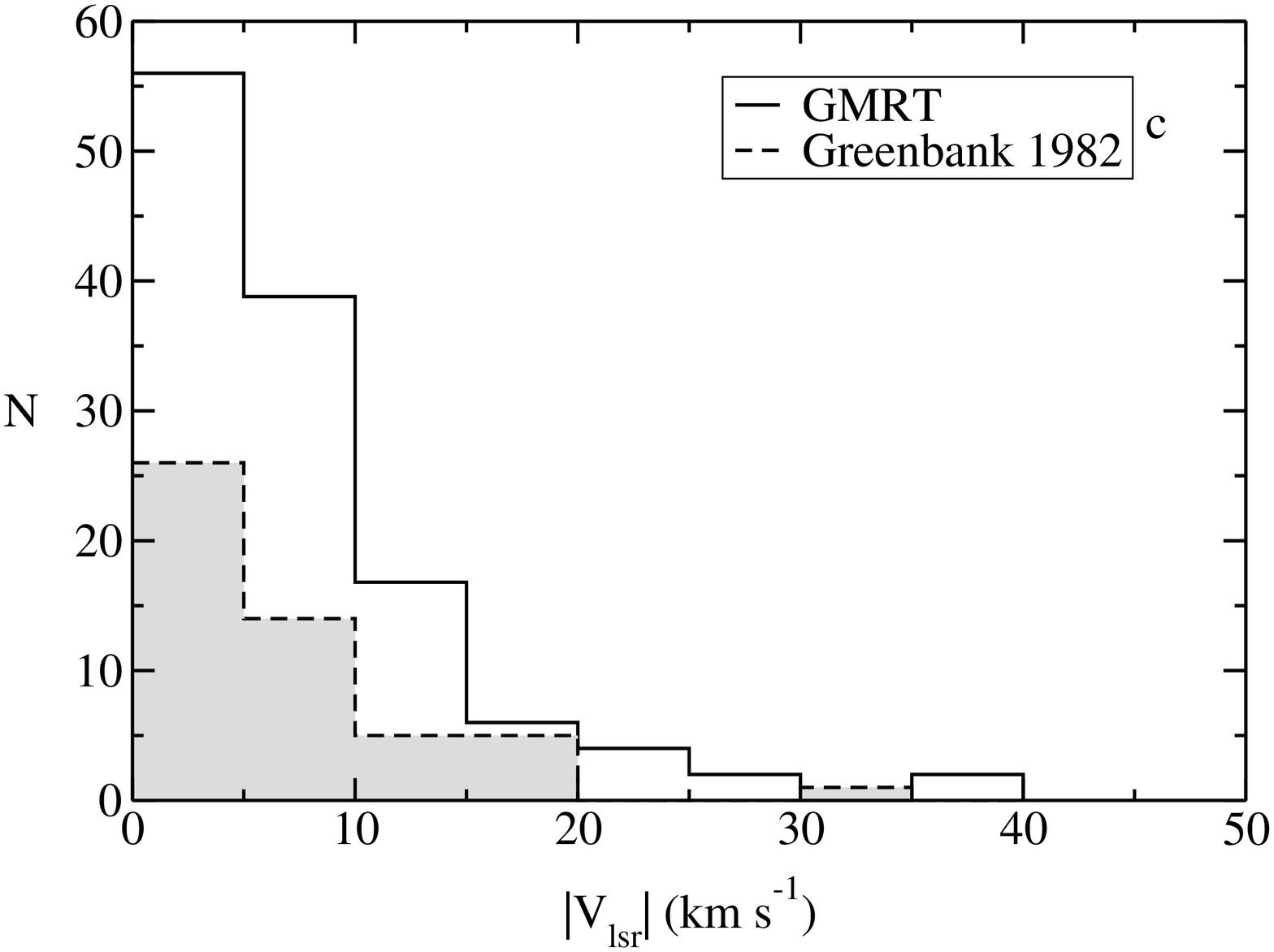}
\includegraphics[scale=0.212,angle=-90]{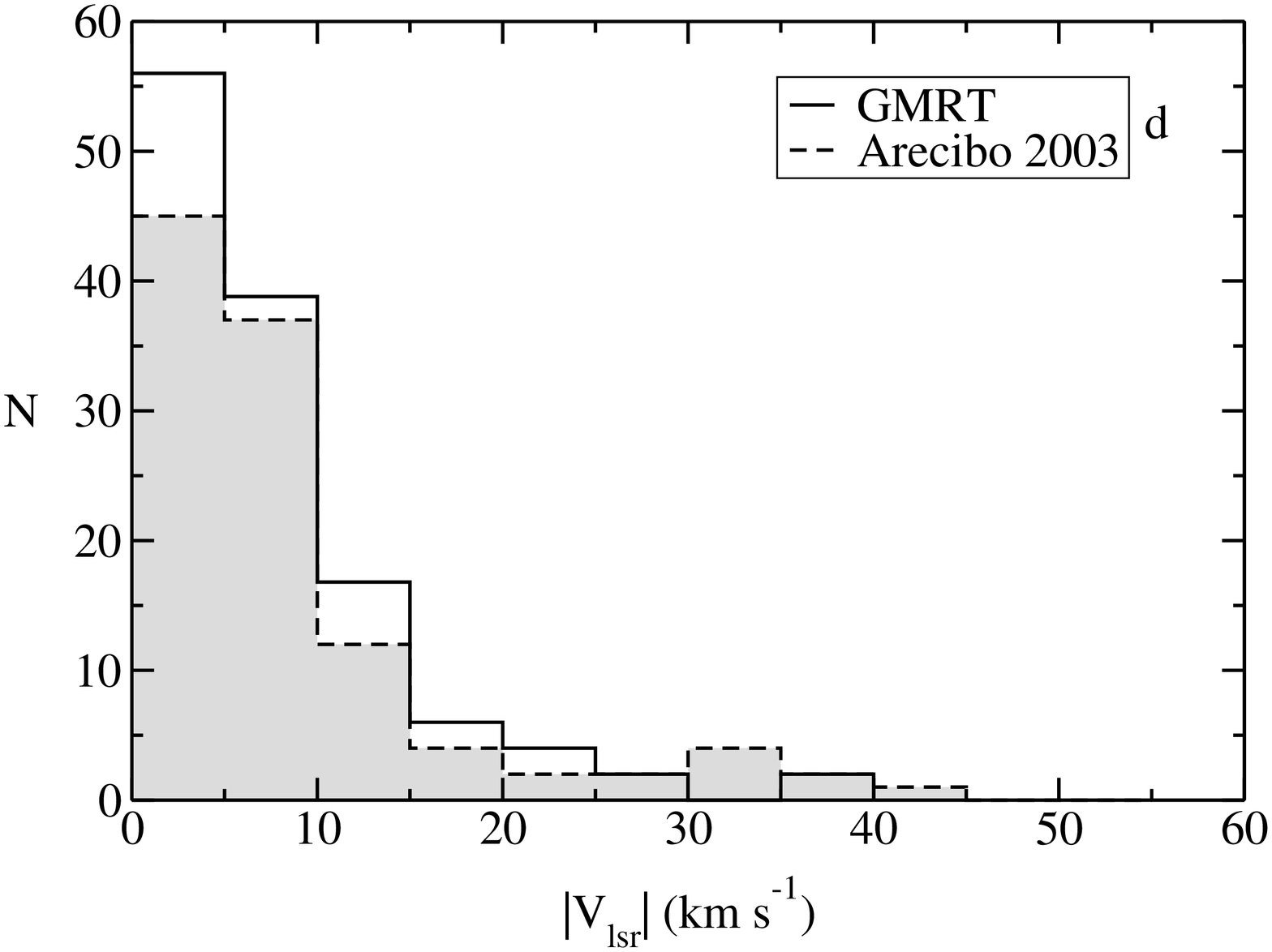}
\end{center}
\caption{The frequency distribution of mean LSR velocities of the
HI absorption features identified from the GMRT survey is compared
with that from Radhakrishnan et al (1972) (a), 
Dickey, Salpeter \& Terzian (1978) (b), Mebold et al 
(1982) (c) and from Heiles \& Troland (2003a) (d) respectively.}
\label{fig:hb_oldies}
\end{figure}

Figure \ref{fig:hb_oldies} shows the frequency distribution of radial
velocities of HI absorption features from the present survey along with that
of 22 HI absorption features from the Parkes Interferometer survey by
Radhakrishnan et al \nocite{rad72b}(1972),  37 discrete HI absorption features
from the Arecibo measurements by Dickey, Salpeter \& Terzian
\nocite{dst78}(1978) and
51 absorption features from the Effelsberg-Green Bank HI absorption survey
by Mebold et al (\nocite{egs81}1981, \nocite{egs82}1982). No feature at
radial velocities larger than 15
km s$^{-1}$ was detected in the Parkes survey (Fig. \ref{fig:hb_oldies}a).
The Arecibo HI absorption survey by Dickey et al. \nocite{dst78}(1978) had an
rms sensitivity comparable to the present survey. Although the number of
absorption features from their survey is smaller, the spread in the
radial velocities are found to be similar to the present survey.
The Effelsberg-Green Bank survey by Mebold et al \nocite{egs82}(1982) lists
69 lines of sight. We have used only the Interferometric measurements from
Mebold et al. \nocite{egs82}(1982). Figure \ref{fig:hb_oldies}c shows a
comparison of radial velocity histogram from the present survey to that
by Mebold et al \nocite{egs82}(1982). It is evident from the figure that 
except for the single absorption feature at $\sim$35 km s$^{-1}$, all the 
features in the survey by Mebold et al \nocite{egs82}(1982) are at
velocities below $\sim$ 20 km s$^{-1}$.

It is clear from figures \ref{fig:hb_oldies}a,b \&c that the present dataset
has not only detected more absorbing clouds, but also more higher velocity
clouds. The peak optical depths of these features at higher velocities
are found to be lower as compared to the rest of the clouds
(See Figure \ref{fig:velotau}). The random velocity distribution of the
features detected in the present survey agrees well with that from the
recent Arecibo survey by Heiles \& Troland (\nocite{ht03a}2003a)
(Fig. \ref{fig:hb_oldies}d).

There are six common sources between the present survey and the Arecibo 
observations. While the rms sensitivity in HI optical depth of the Arecibo
survey is slightly worse ($\tau_{\scriptscriptstyle{\rm{HI}}}$ $\sim$0.006)
compared to the present survey, its velocity
resolution (0.16 km s$^{-1}$) is better compared to the current
 HI absorption measurements
(3.3 km s$^{-1}$). Therefore, the number of HI absorption components detected
in their observations are higher than that in the present survey. 
For the six common sources, the fitted values for the peak optical depths 
from the present survey and the Arecibo survey agree to within $\sim$ 30\%.
The fitted values for line centers and widths of the corresponding features
differ by less than $\sim$ 2 km s$^{-1}$ between the two surveys.

\subsection{Comparison of the GMRT HI Absorption data with the Optical surveys}
\label{sec:highb_opt}
The CaII and NaI absorption studies of Adams \nocite{adams}(1949),
Blaauw \nocite{b52}(1952), M\"unch \nocite{m57}(1957), M\"unch \& Zirin
\nocite{m61}(1961) and others revealed two set of
absorption features, one at lower and the
other at higher random velocities. More recently, high resolution spectra
of NaI lines
(eg. Welty et al, \nocite{whk94}1994) and CaII lines (Welty et al,
\nocite{wmh96}1996) were
obtained towards a number of stars. 

\begin{figure}
\begin{center}
\includegraphics[width=5.7cm,height=10.5cm,angle=-90]{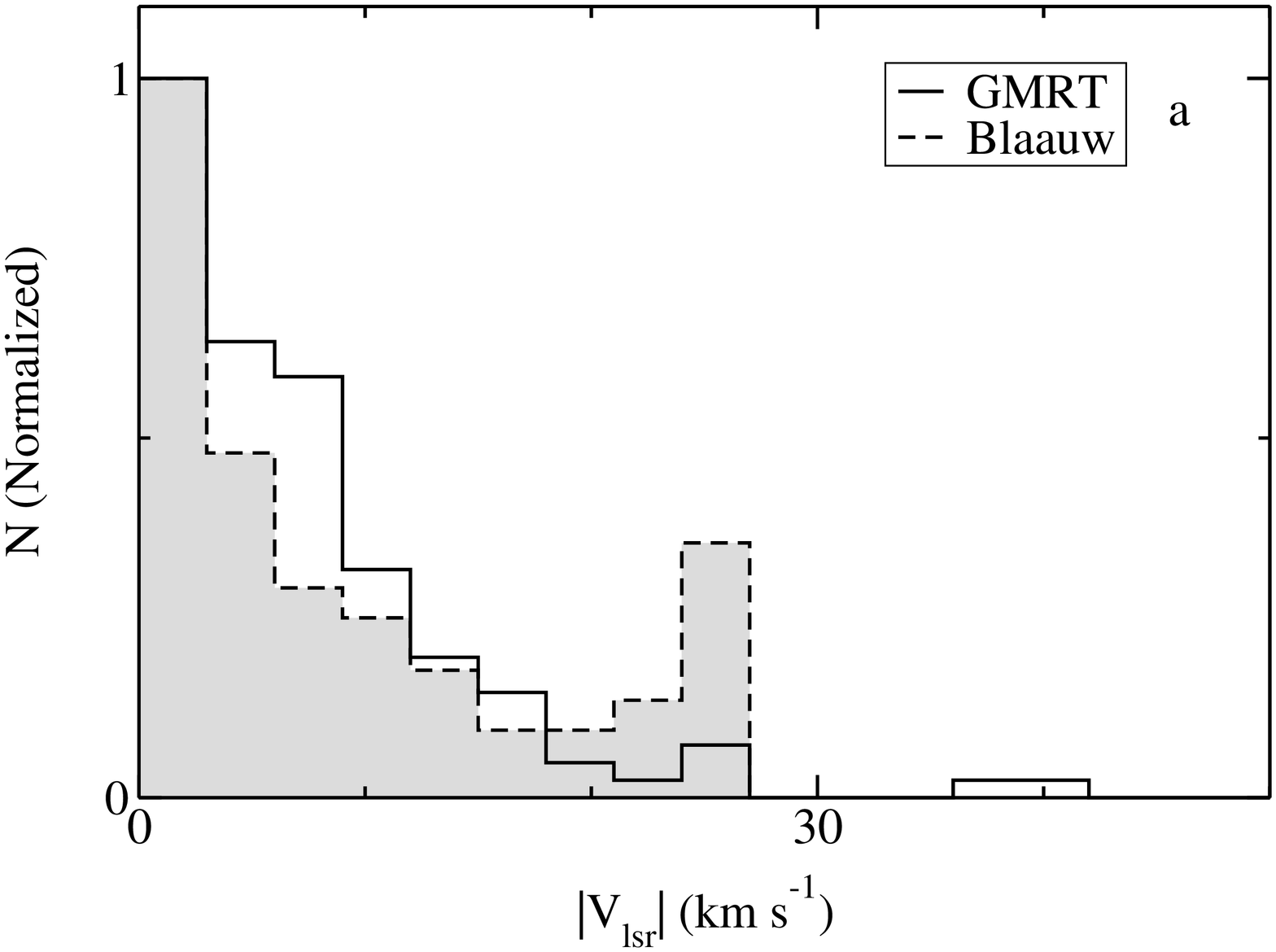}
\includegraphics[width=5.7cm,height=10.5cm,angle=-90]{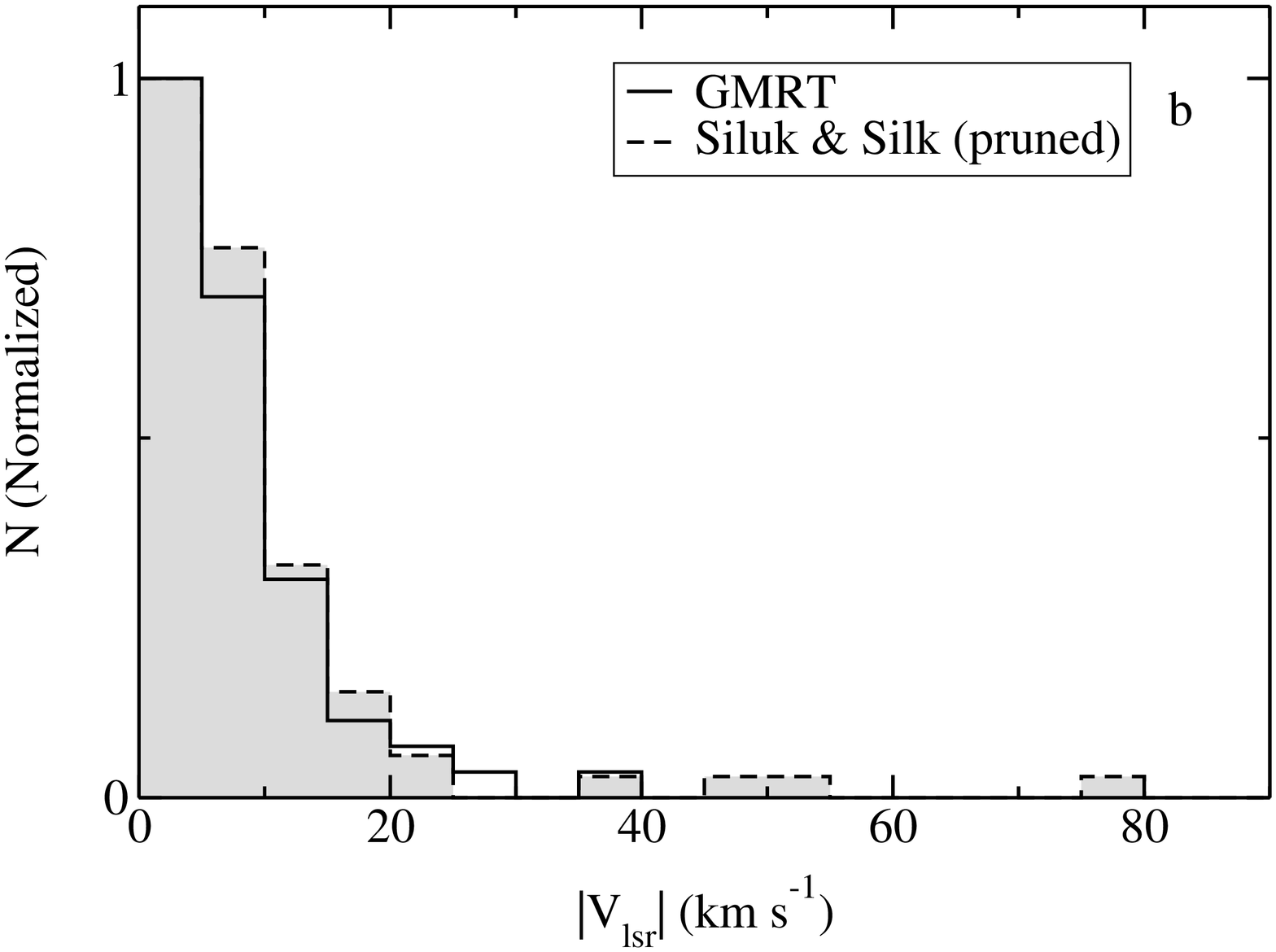}
\includegraphics[width=5.7cm,height=10.5cm,angle=-90]{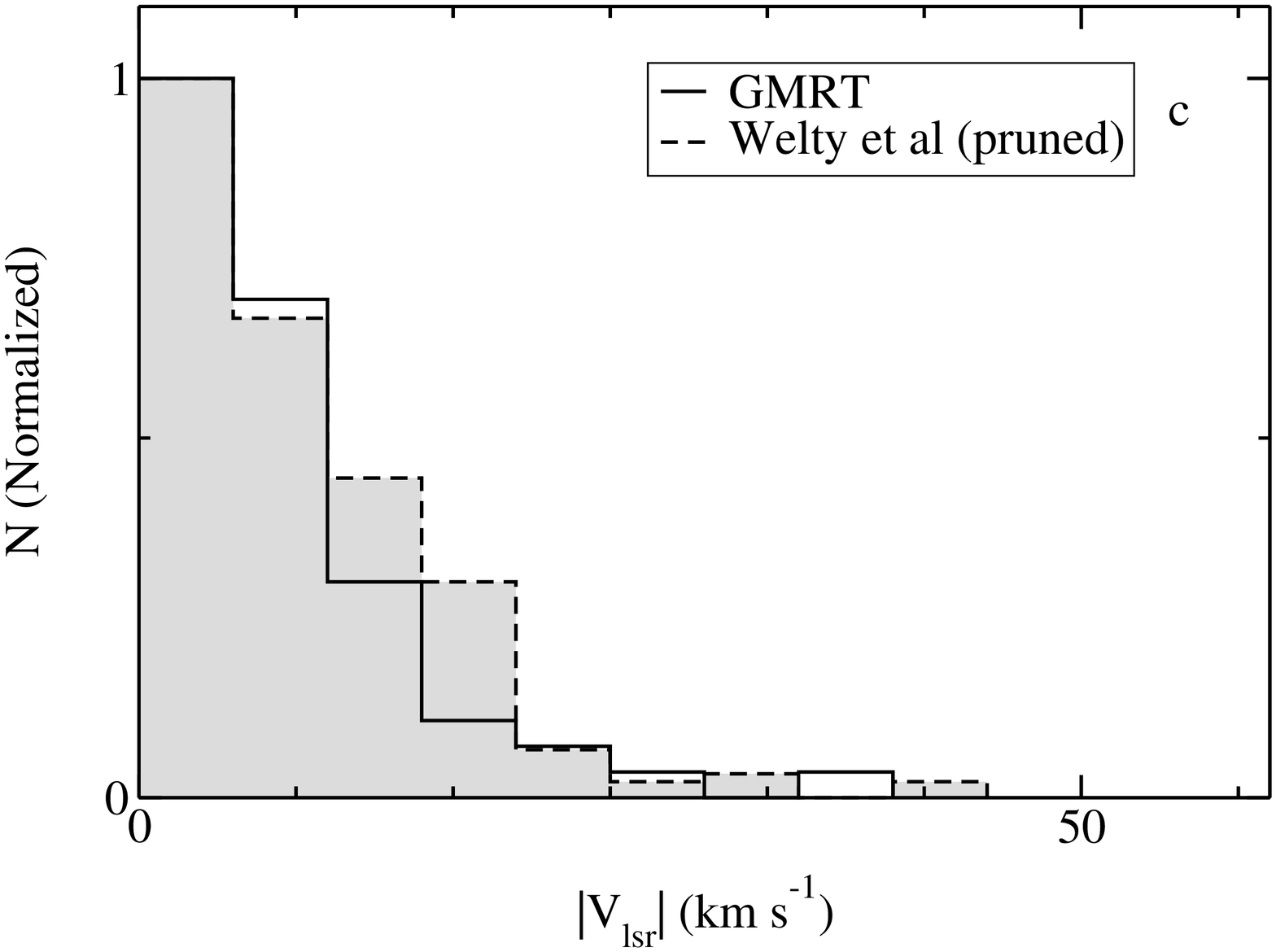}
\end{center}
\caption{The frequency distribution of the radial velocities
of HI absorption features identified from the present survey along with those
from optical surveys. (a): Blaauw, 1952 (dashed line).
The secondary peak at high velocity in Blaauw's histogram is an artifact of the
binning used by Blaauw. (b): Optical absorption
line data from Siluk \& Silk, 1974 (pruned), and (c):
The CaII absorption survey by Welty, Morton \& Hobbs, 1996
(pruned).}
\label{fig:hb_opt}
\end{figure}

Figure \ref{fig:hb_opt}a shows the frequency distribution of the
velocities of the HI absorption features from the present survey with that
from Blaauw's \nocite{b52}(1952) study. The secondary peak at
high velocity in Blaauw's histogram is an artifact of the binning used by
Blaauw \nocite{b52}(1952). He counted all the features with radial 
velocities V$_{lsr}$ $\ge$ 21 km s$^{-1}$ in a single bin. Excluding
this secondary peak, there is reasonable agreement between the two
histograms.

In Fig. 4b, we compare the velocity distribution of HI absorption components
from the present survey
with that from Siluk \& Silk \nocite{ss74}(1974).
Some of the absorption at higher velocities in their data may not represent
random velocities of diffuse HI clouds; they could be due to large
systematic motions (Spiral arms, IVCs, HVCs, etc.) superimposed on random
motions (Mohan, \nocite{me2003}2003).
Excluding those features, we have re-constructed the histogram
of optical absorption line velocities from the data of Siluk \& Silk (Fig
\ref{fig:hb_opt}b). The modified histogram from Siluk \& Silk 
\nocite{ss74}(1974) is in good agreement with the current HI absorption 
velocity histogram. 

Fig. \ref{fig:hb_opt}c compares the frequency distributions 
of radial velocities from the present survey with that of a recent high 
resolution survey of CaII lines (Welty et al. \nocite{wmh96}1996). 
Out of the 44
stars towards which they measured CaII absorption, eight stars were located at
a distance of 450 pc, in the constellation Orion, and one star, HD72127, in
Vela. The CaII absorption features
towards these nine stars were not included in the comparison since the velocities
of the absorption features were dominated by systematic motions.

It is clear from figure \ref{fig:hb_opt} that the extent of velocities of CaII
absorption line features are comparable to that of the HI absorption lines in
the present survey. The problem of larger spread in the 
velocities of interstellar optical absorption lines as compared to the HI 
21-cm line features have been an open problem for decades. Our analysis makes 
it clear that many of the often quoted interstellar optical absorption 
features at higher velocities may not represent true random motions. Leaving 
aside such features, the higher velocity HI absorption features detected in 
the present survey can account for the ``missing high random velocity" 
interstellar clouds in the earlier HI 21-cm line studies.

\section{Statistics of the HI absorption line parameters}
\label{sec:abs}
The details of the observing strategy, the HI absorption spectra from GMRT
and the discrete line components identified by fitting Gaussians to
the spectra are presented in \nocite{me2004}paper I. Here we discuss 
their interpretation.

\subsection{The frequency distribution of HI absorption line parameters}
\label{sec:lineparams}
\begin{figure}
\begin{center}
\includegraphics[width=5.7cm,height=9.5cm,angle=-90]{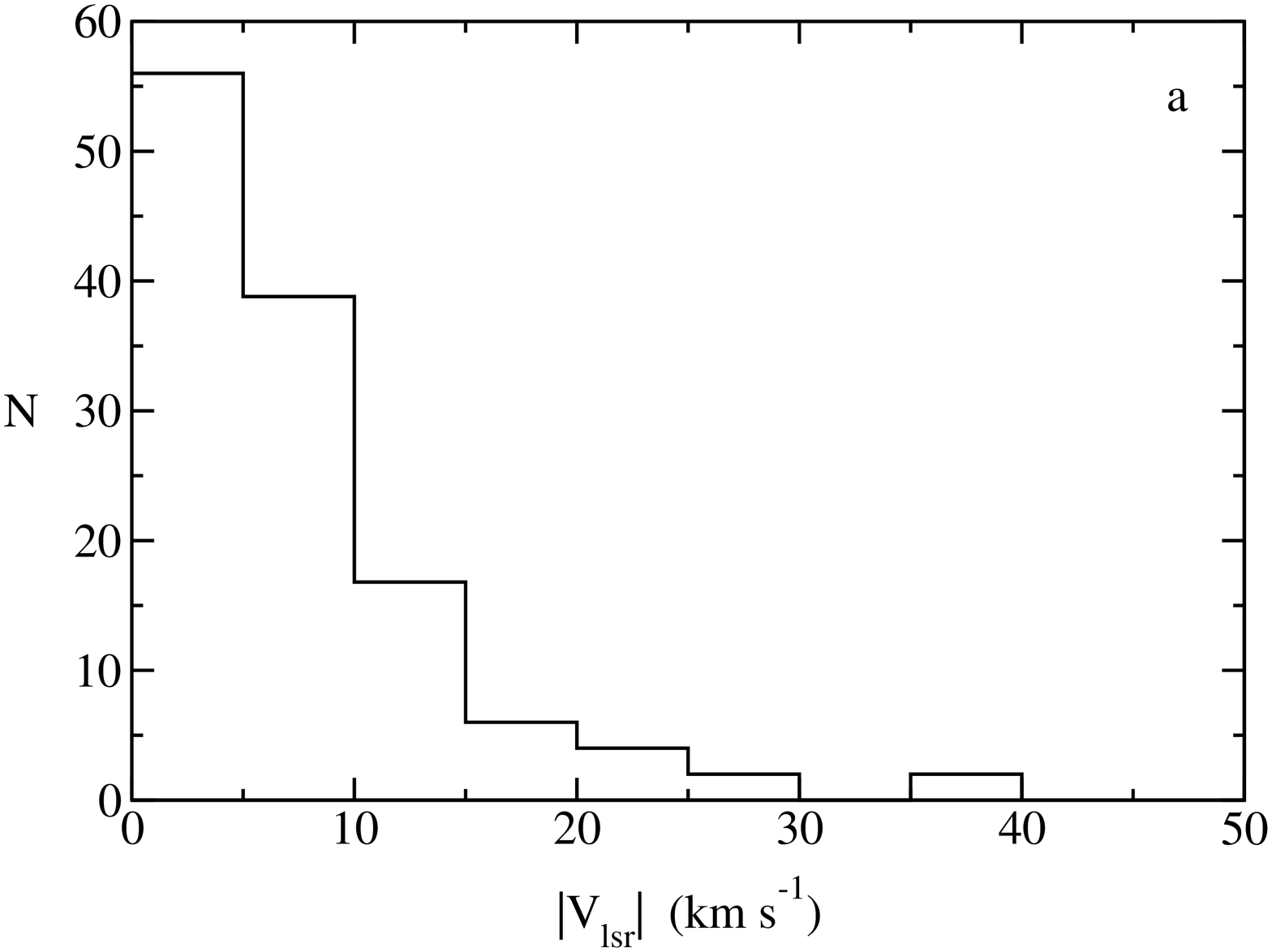}
\includegraphics[width=5.7cm,height=9.5cm,angle=-90]{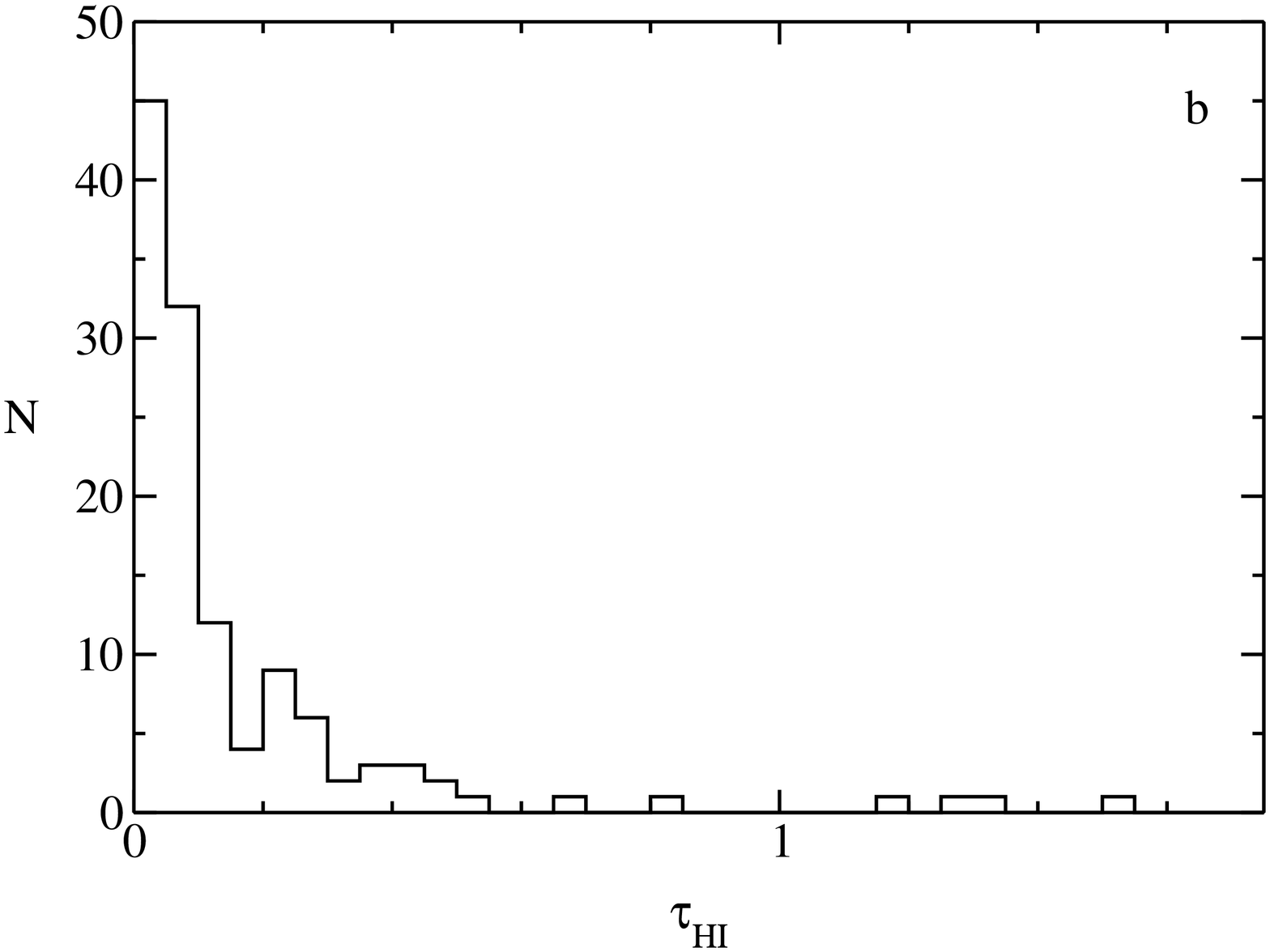}
\includegraphics[width=5.7cm,height=9.5cm,angle=-90]{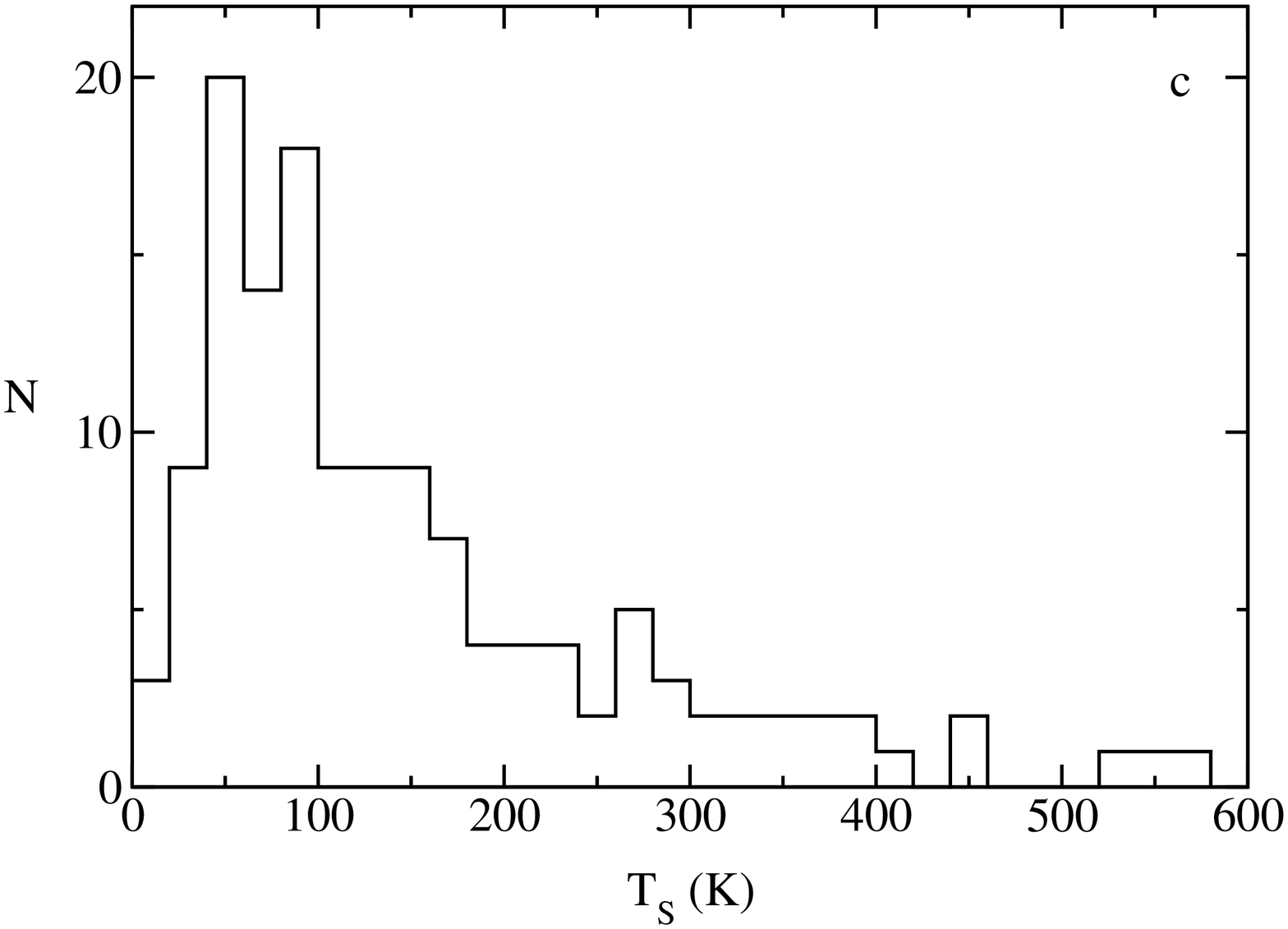}
\end{center}
\caption{The frequency distribution of the absorption line parameters from the 
present survey. (a): The histogram of random velocities of the HI absorption 
components, in 5 km s$^{-1}$ bins. (b): The frequency distribution of peak 
optical depths of HI absorption features, in bins of 0.05 and (c): The 
frequency distribution of estimated spin temperatures of the HI absorption 
features in 20 K bins.}
\label{fig:lineparams}
\end{figure}

The frequency distribution of the mean LSR velocities of the HI absorption line
components is shown in figure \ref{fig:lineparams}a. 
This histogram was discussed in section \ref{sec:old_radio+opt}.
It was noticed in earlier studies that the average optical depth 
of HI  absorption features is higher
at lower Galactic latitudes (See for eg. Mebold et al., \nocite{egs82}1982). 
This apparent
excess of large optical depth features in the Galactic plane has been
attributed to superposition of absorbing clouds along the line of sight.
However, there are only seven features in our survey with their peak optical 
depths above 0.5 (Fig. \ref{fig:lineparams}b). This is an evidence that in the
present dataset the superposition of more than one HI absorption feature along
the line of sight is minimal. This is expected since we are sampling
relatively small path lengths through the gas layer at higher Galactic
latitudes.

As we discussed in paper I, we have used the HI emission data from the 
Leiden-Dwingeloo survey along with the present GMRT HI absorption data to
estimate the spin temperatures of the HI absorption features. 
The mean value of the spin temperature from the present survey was found
to be $\sim$120K, with considerable scatter. For the lower optical depth 
features ($\tau$ $\ltsim$ 0.1), 
we find spin temperature, $T_{\scriptscriptstyle{\rm{S}}}$ = 150 $\pm$ 78 K. 
For the higher optical depth features ($\tau$ $\gtsim$ 0.1), we find
$T_{\scriptscriptstyle{\rm{S}}}$ = 74 $\pm$ 30 K. The mean spin temperature of
the lower optical depth features, though has a large scatter, is two times 
higher than that of the higher optical depth features.
Figure \ref{fig:lineparams}c shows the histogram of  the spin temperatures of 
the HI absorption features. The distribution of spin temperature
peaks below 100 K, which agrees in particular with Heiles 
\& Troland (\nocite{ht03b}2003b), who found the spin temperature distribution 
to peak near 40K. Both the Arecibo observations and the present survey 
indicate the presence of a higher temperature ``tail'' in the spin temperature
histogram. 

\subsection{The $\tau -$ 
T$_{\scriptscriptstyle{\rm{S}}}$ relation}
\label{sec:tauts}
\begin{figure}
\begin{center}
\includegraphics[width=6.0cm,height=9.5cm,angle=-90]{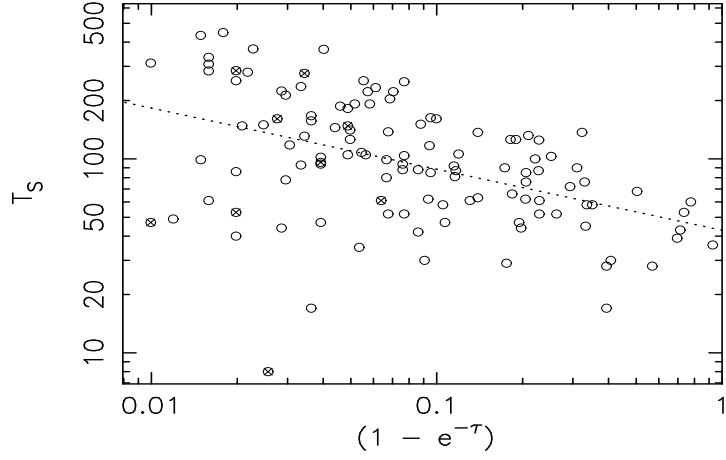}\\
\end{center}
\caption{log(1 - e$^{-\tau}$) $-$ log(T$_{\scriptscriptstyle{\rm{S}}}$) 
plot for the HI absorption features detected in the present survey. HI emission
data are from the Leiden-Dwingeloo survey. The dotted line is the best fit
for the observed data points. The circles with crosses inside them correspond 
to the high random velocity HI absorption features (Table 1).}
\label{fig:tauts}
\end{figure}

Several studies in the past had reckoned an inverse correlation between
log(1 - e$^{-\tau}$) and log(T$_{\scriptscriptstyle{\rm{S}}}$)
(Lazareff, \nocite{laz75}1975; Dickey et al., \nocite{dst79}1979; 
Crovisier \nocite{crow81}1981). It was noted that a relationship
of the form

\begin{equation}
log\-T_{\scriptscriptstyle{\rm{S}}} = log\-T_{\scriptscriptstyle{\rm{S}0}} + A\-log(1 - e^{-\tau})
\label{eqn:tauts}
\end{equation}
\noindent
exists between the observed optical depth and the estimated spin temperature.
However, Mebold et al (\nocite{egs82}1982) found no significant correlation 
between the spin temperature and optical depth.
Figure
\ref{fig:tauts} is a plot of (1 - e$^{-\tau}$) 
$v/s$ (T$_{\scriptscriptstyle{\rm{S}}}$) from the present dataset.
We do find an inverse correlation, though the scatter is larger
at lower optical depths.
A least square fit to the present dataset provides
$T_{\scriptscriptstyle{\rm{S}0}}$ = 43K and $A$ = --0.31 (eqn \ref{eqn:tauts}). 
The values obtained from the previous studies were $\sim$60K and --0.35
respectively (Kulkarni \& Heiles, \nocite{sr88}1988).
A similar least-squares fit to the recent Arecibo data yields
$A$ = -- 0.29 and T$_{\scriptscriptstyle{\rm{S}0}}$ = 33K.
However, Heiles \& Troland (\nocite{ht03b}2003b) have carried out a 
detailed analysis of the correlation between the various parameters
like the spin temperature, the HI column density, the optical depth
and the kinetic temperature of the absorption features detected in the
Arecibo survey. They emphasize that the mutual correlation that exists between the
four parameters renders meaningless the results of least-square fit carried out
on only selected pairs of variables. They conclude that there is no physically
significant relation between optical depth, spin temperature and HI column
density.

\subsection{The $\tau_{\scriptscriptstyle{\rm{HI}}}$ $-$ 
V$_{\scriptscriptstyle{\rm{LSR}}}$ relation}
\label{sec:velotau}
\begin{figure}
\begin{center}
\includegraphics[width=6.0cm,height=9.5cm,angle=-90]{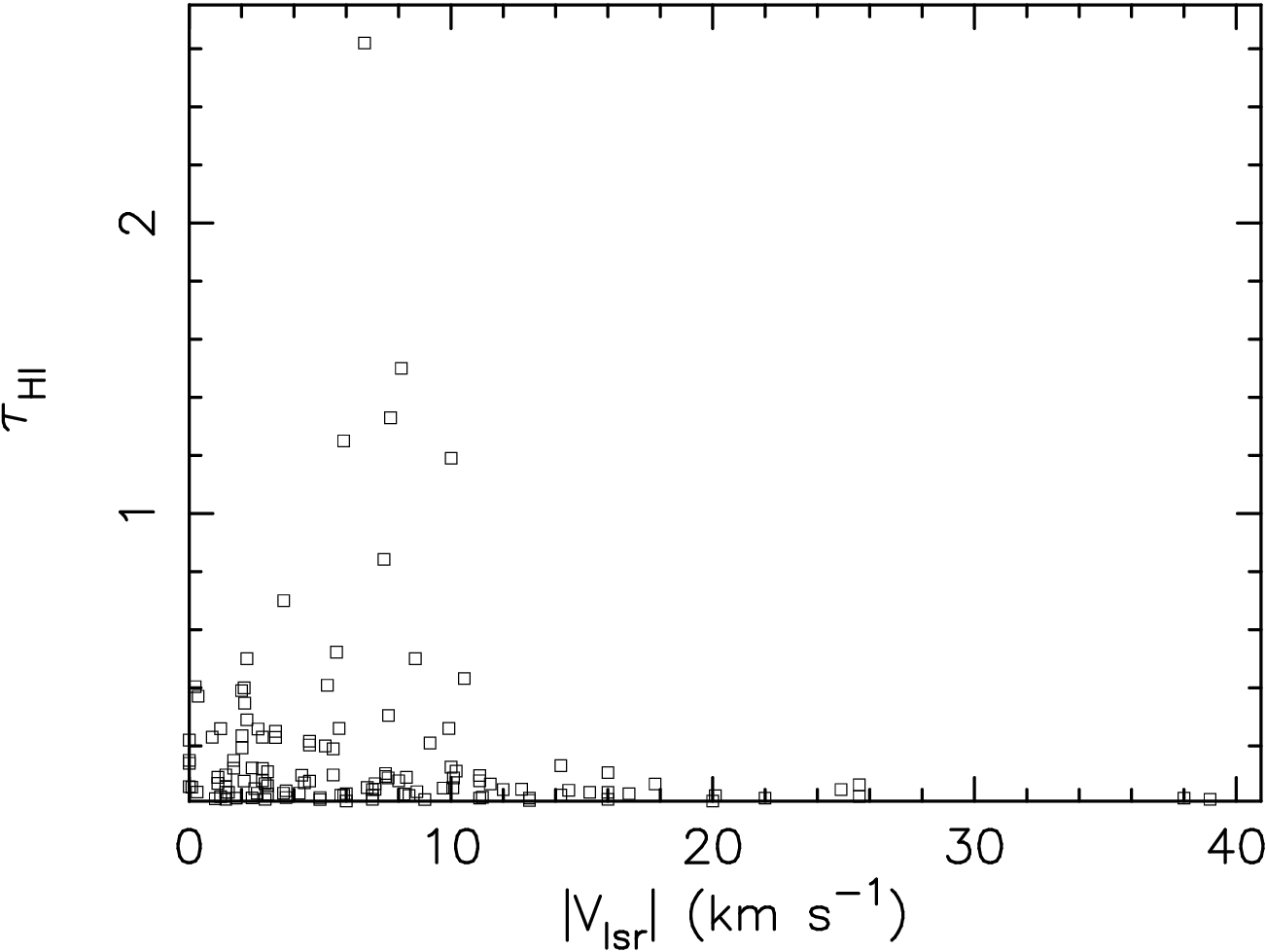}\\
\end{center}
\caption{$\tau_{\scriptscriptstyle{\rm{HI}}}$ $-$ 
$|V_{\scriptscriptstyle{\rm{LSR}}}|$ plot. The higher optical depth features
are confined to $|V_{\scriptscriptstyle{\rm{LSR}}}|$ $\ltsim$ 15 km s$^{-1}$.}
\label{fig:velotau}
\end{figure}

The peak
optical depth as a function of $|V_{\scriptscriptstyle{\rm{LSR}}}|$ is
shown in figure \ref{fig:velotau}. The peak
$\tau_{\scriptscriptstyle{\rm{HI}}}$ drops sharply near
$|V_{\scriptscriptstyle{\rm{lsr}}}|$ $\sim$ 10 km s$^{-1}$.
No spectral features at velocities above 10 km s$^{-1}$ have peak optical depth
above $\sim$ 0.1. On the other hand, low optical depth features are detected 
over a larger velocity range. In section \ref{sec:lineparams} we found that 
lower optical depth features have higher mean spin temperature, almost a factor 
of two higher than the higher optical depth features. Fig. \ref{fig:velotau} 
conveys an extra information that these higher spin temperature features are 
also spread over a larger range in random velocity.

\subsection{The distribution of the HI optical depth as a function of $b$}
\label{sec:tau_b}

\begin{figure}
\begin{center}
\includegraphics[width=6.0cm,height=9.5cm,angle=-90]{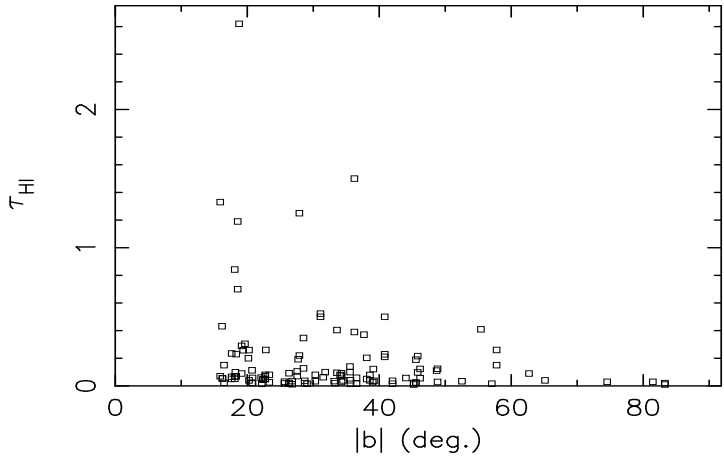}
\end{center}
\caption{The distribution of optical depth of the HI absorption line features
as a function of Galactic latitude. The number of higher optical depth features
drop gradually with increasing latitude, but the number of low optical depth
features ($\tau$ $\ltsim$ 0.1) remain more or less the same.}
\label{fig:tau_b}
\end{figure}

HI optical depth of the absorption features are plotted against their
Galactic latitude in Figure \ref{fig:tau_b}. The optically thick components
($\tau$ $>$ 1.0), though
very few, are all confined to $|b|$ $\ltsim$ 30$^{\circ}$.
The low optical depth features ($\tau$ $\ltsim$ 0.1), on the other
hand, are almost uniformly distributed with respect to the Galactic latitude.
This is an indication that the scale height of the low optical depth features
are different from that of the higher optical depth features.

\subsection{The velocity dispersion of interstellar Clouds}
\label{sec:histofitsec}
\begin{figure}
\begin{center}
\includegraphics[width=6.0cm,height=10.0cm,angle=-90]{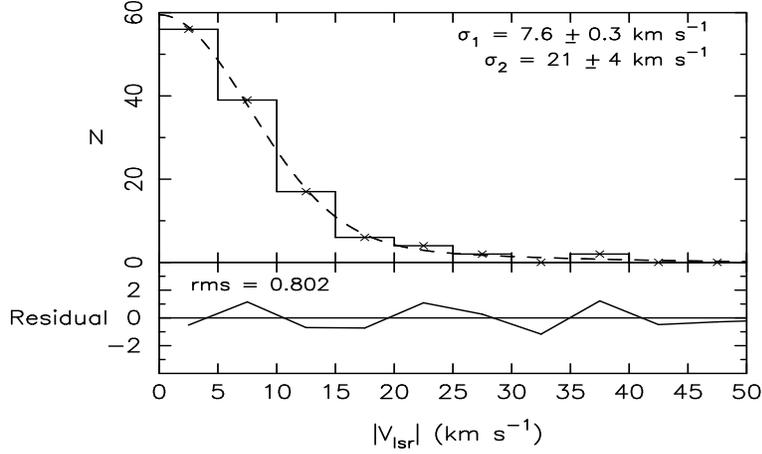}
\end{center}
\caption{The frequency distribution of LSR (random) velocities of HI absorption
components from the present survey. The dashed line is the sum of two
Gaussian components with $\sigma_1$ = 7.6 $\pm$ 0.3 km s$^{-1}$ and
$\sigma_2$ = 21 $\pm$ 4 km s$^{-1}$. Both the Gaussians are centered at
v$_{\scriptscriptstyle{\rm{LSR}}}$ = 0 km s$^{-1}$. The lower panel shows 
the residual after subtracting the fitted model from the observed 
distribution.}
\label{fig:tau_histo_gfit}
\end{figure}

The velocity distribution of cold atomic gas in the Galaxy is known to be
a Gaussian with a dispersion of $\sim$7 km s$^{-1}$ (Dickey \& Lockman,
\nocite{araa90}1990).
As we concluded in section \ref{sec:galrot}, the observed radial velocities of the
HI absorption components detected in the present survey are essentially random
velocities.
We have carried out a Gaussian fit to the observed frequency distribution of
the radial velocities of the HI absorption features detected in the present study
(Fig. \ref{fig:tau_histo_gfit}). A non-linear least square method was used to
fit Gaussians to this  histogram. A two Gaussian model was found to be a
good fit for the distribution, with a reduced chi-square value of 1.4.
The low velocity features were found to form a Gaussian distribution with
a velocity dispersion of $\sigma_{1}$
 = 7.6 $\pm$ 0.3 km s$^{-1}$, which agrees well with the earlier results.
For eg., Belfort \& Crovisier \nocite{bcrov84}(1984) had performed a
statistical analysis 
of the radial velocities of HI clouds observed by surveys using the Arecibo
(Dickey et al, \nocite{dst78}1978), and Effelsberg \& Green Bank
(Mebold et al, \nocite{egs82}1982).
Their value for the velocity dispersion of HI clouds was $\sim$6.9 km s$^{-1}$. 
The higher
velocity features in the present survey seem to form a distribution with
a velocity dispersion of $\sigma_{2}$ = 21 $\pm$ 4 km s$^{-1}$
(Fig. \ref{fig:tau_histo_gfit}). The presence of two Gaussian features in the
velocity distribution of interstellar clouds is indicative of two distinct
populations of interstellar clouds.
From the area under the respective curves,
this data indicates that $\sim$20\% of the clouds belong to the second
population, with larger velocity dispersion.

A study of the velocities of interstellar optical absorption lines
(NaI \& CaII) were carried out by Sembach \& Danks \nocite{sd94}(1994).
They observed that only a two component model can be fitted to the velocity
distribution of CaII line components. The values which they obtained were:
$\sigma_1$ $\sim$8 km s$^{-1}$ and $\sigma_2$ $\sim$21 km s$^{-1}$, which
agree well with the velocity dispersions of the two components obtained
from the present data of HI absorption lines.

\subsection{The High velocity HI absorption features}
\label{sec:highvelo}
We have detected 13 HI absorption features at higher random velocities
($|v|$ $>$ 15 km s$^{-1}$), out of the 126 total HI absorption features
(Table \ref{tab:highvelo}).
The optical depths
of these higher velocity features are below 0.1. The mean value of
peak optical depth of all these features is
0.04 $\pm$ 0.02. In most cases, we could identify the corresponding HI 
emission feature in the HI emission profile from the Leiden-Dwingeloo survey.
If the velocity difference between the absorption and the emission
line was less than or comparable to the channel width of the GMRT
observations ($\sim$3.3 km s$^{-1}$), and if the difference between line widths
was within $\sim$ 5 km s$^{-1}$, the
spectral lines were assumed to originate in the same physical feature. The mean
brightness temperature of these features was $\sim$4 K and a mean HI column
density of (4.3 $\pm$ 3.4) $\times$ 10$^{19}$ cm$^{-2}$. The mean value of the spin
temperatures of these clouds is 125 $\pm$ 82 K. 

\begin{table}
\caption{The higher random velocity ($|v|$ $\gtsim$ 15 km s$^{-1}$)
HI absorption features detected in the present survey.
Columns 2, 3 and 4 list the peak optical depth, the
mean LSR velocity and the FWHM respectively of discrete components identified
using the Gaussian fitting.
The value of FWHM is deconvolved for a channel width of 3.26 km s$^{-1}$. The
unresolved lines are marked with a "--" . Columns 5, 6 and 7 list the
same for the HI emission profile along the same line of sight, obtained from the
Leiden-Dwingeloo survey of Galactic neutral hydrogen (Hartman \& Burton, 
1995)
The formal 1$\sigma$ errors estimated in the last digit of the fitted
parameters are given within brackets. The implied HI column densities are
listed in column 8 and the estimated spin temperatures are given in column 9.}
\label{tab:highvelo}
\begin{center}
\begin{small}
\begin{tabular}{l l l l l l l l l l} \hline \hline
         & \multicolumn{3}{c}{{\bf HI Absorption} (GMRT)} & & \multicolumn{4}{c}{{\bf HI Emission} (LDS)}    &   \\ \cline{2-4} \cline{6-9}
         &         &     &             &             &     &      &             &      &              \\
Source   &$\tau_{\scriptscriptstyle{\rm{HI}}}$  &  v$_{\scriptscriptstyle{\rm{lsr}}}$  &    $\Delta$v  &    & T$_{\scriptscriptstyle{\rm{B}}}$    & v$_{\scriptscriptstyle{\rm{lsr}}}$   & $\Delta$v & N$_{\scriptscriptstyle{\rm{HI}}}$ & T$_{\scriptscriptstyle{\rm{S}}}$   \\
         &         &     &             &             &     &      &             & $\times$ 10$^{19}$     &              \\
         &             &(km/s)&(km/s)&&  (K)      &(km/s)&(km/s)& cm$^{-2}$ & (K)   \\ \hline
J0459+024& 0.026(4) &--25.6(5)  &4.6(8) &    &0.21(8)&--28(1)      &6(3)  & 0.24  &  8 \\
J0541--056 & 0.069(7) &+17.8(9)   & 3.2(5)  &  &  ---     &   ---       & --- & ---   &   \\
J0814+459 &  0.035(2) & +16.8(9)  & 4(2)   &     &9.5(5)  &+15.4(1)   &3.8(1)& 7.0 &    276        \\
J1154--350 &  0.016(3) &--16(2)      &13(5) & & ---   &  ---        &  ---   & ---   &    \\
J1221+282 & $\sim$0.01   &--20         & -- &         &0.47(4)  &--18.9(4)    &11(1)   & 1.0   &  47         \\
&  $\sim$0.02   &--38      &    --  &        &1.06(4)  &--34.8(1)    &6.1(3)   & 1.2 & 53         \\
J1257--319 & 0.108(6)&--16.0(2)  &3.8(3) &    & ---        & ---           &   ---        & --- &         \\
J1351--148 &  0.020(4)& +22.0(6)  &5(1)  &       &  5.7(1)  & +22.9(1)    & 10.5(2)    &  12.0 & 285      \\
J1554--270 &  0.04(2)   &--16(8)      & 8(5)  &      &  ---       &  ---          &    ---  & ---     &        \\
J1638+625 &  0.028(4) & --20.1(4) & 5(1)   &     & 4.50(6) & --21.05(2) & 5.05(7) & 4.4  & 161        \\
J1751+096 &  0.066(3) &+25.6(2)   &4.6(2)  &   &3.9(1)    &+25.1(1)     &7.1(3) &  5.3  &  61         \\
J2005+778 &  0.05(1)   &--25(4)  & 2(2)    &    &7.2(3)    &--23.59(8)  &7.4(2) & 10.0  & 148        \\
J2232+117 &  0.040(4) & --15.3(3) & 2.9(7) &   & 3.83(7) & --13.88(3) & 3.92(8) & 2.3 &  96         \\ \hline \hline
\end{tabular}
\end{small}
\end{center}
\end{table}

We analysed the HI emission profiles from the Leiden-Dwingeloo survey
towards 91 directions along which we measured HI absorption using the GMRT.
HI column density of individual features along these directions were
calculated using the emission data.
However, 11 out of the 102 directions studied using the GMRT were beyond the
declination limit of the Leiden-Dwingeloo survey. Towards these sources we
assumed a spin temperature of 120 K and 
the observed widths of the HI absorption lines to estimate the
column density.

\begin{figure}
\begin{center}
\includegraphics[width=7.0cm,height=10.0cm,angle=-90]{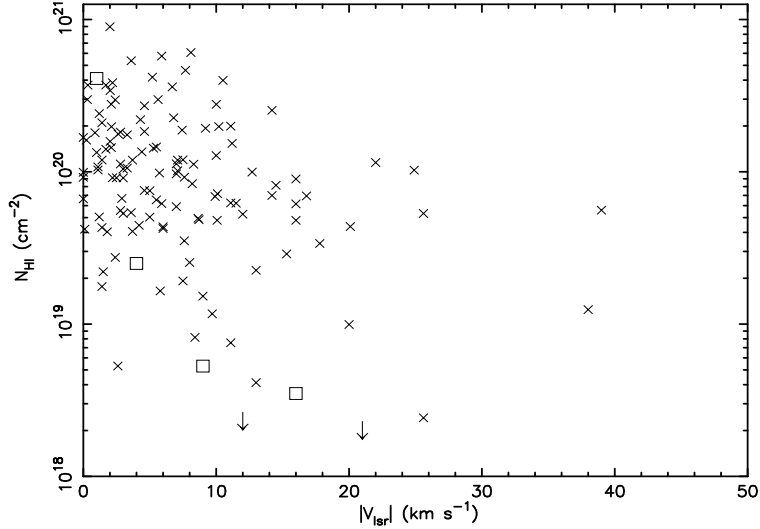}
\end{center}
\caption{The column densities of the HI absorption features detected in the
present survey. The brightness temperature and line width of the HI emission
feature corresponding to each absorption feature was obtained from the
Leiden-Dwingeloo survey. For those directions where HI emission data is not
available, we have assumed a spin temperature of 120K and the velocity width
of the HI absorption line to calculate the column density.
The squares are the HI column density estimates
from the UV absorption line data from Martin and York (1982).
and the arrows are the upper limits.}
\label{fig:nh_v_plt}
\end{figure}

The estimated HI column density
as a function of the LSR velocity is shown in fig. \ref{fig:nh_v_plt}.
The measured HI column density ranges from
$\sim$2.4 $\times$ 10$^{18}$ to $\sim$9 $\times$ $^{20}$ cm$^{-2}$.
It is clear from fig. \ref{fig:nh_v_plt} that the
column densities of the features are decreasing systematically with increasing
velocity. This is in agreement with the results from UV absorption line
studies (Martin \& York, \nocite{uv82}1982; Hobbs, \nocite{hobbs84}1984).

\section{Discussion}
\label{sec:discuss}
The present high-latitude, high-sensitivity HI absorption survey from
the GMRT has confirmed many of the known results. These results are
summarised in the histograms (Fig. \ref{fig:lineparams}) of random velocity, 
peak optical depth, and spin temperature of the absorption components. 
In addition, variation
of peak optical depth w.r.t. spin temperature, random velocity
and Galactic latitude are displayed in Figs. \ref{fig:tauts},
\ref{fig:velotau} and \ref{fig:tau_b} respectively. However, a completely new result
is the detection of the low optical depth absorption features forming 
the high velocity tail in the histogram shown in Fig. \ref{fig:lineparams}a. 
Most of these low optical depth features are below the detection limit of 
the earlier HI absorption surveys. The velocity histogram 
(Fig. \ref{fig:lineparams}a) is well-fit by two Gaussians indicating two 
populations of HI absorbing clouds identified by velocity dispersions 
$\sim$ 7 and $\sim$ 21 km s$^{-1}$ respectively (Fig. \ref{fig:tau_histo_gfit}). 
While the slow clouds were first detected in the optical
absorption lines and subsequently in the HI absorption and emission 
surveys, the fast clouds were only detected in optical absorption lines. 
The non-detection of the fast clouds remained a puzzle for a long time. 
The present observations have clarified the nature of the fast clouds to 
some extent.

There have been attempts in the past to explain the high velocity tail
seen in the histogram of random velocities of the optical absorption
lines ( Siluk \& Silk \nocite{ss74}(1974), Radhakrishnan \& Srinivasan
\nocite{rs80}(1980), Rajagopal et al \nocite{jd98b}(1998)).
According to these authors, the high velocity optical absorption lines 
arise in interstellar clouds, shocked and
accelerated by supernova remnants in their late phases of evolution.
The fast clouds are therefore warmer
and also of lower HI column density as compared to the slow clouds, due
to shock heating and evaporation.  The present observations
indicate that these fast clouds have three times
larger velocity dispersion and ten times lower column
densities compared to the slow clouds as might be expected if they
were from a shocked population of clouds.
The decrease in the HI column densities of the fast clouds as a 
function of their random velocities (Fig. \ref{fig:nh_v_plt}) 
is also consistent 
with this scenario. The shocked HI clouds are also
expected to be warmer than the slow clouds. The mean spin temperature 
of the fast clouds detected in the GMRT survey is similar to 
that of the standard slow clouds (Fig. 6). However,
this might be a selection effect since for a given optical depth 
detection limit and an HI column density, clouds with lower spin 
temperature will be preferentially detected.

The fast clouds with three times higher dispersion are expected to have 
a scale height
about ten times larger compared to the slow HI clouds.
Given an effective thickness of 250 pc for the slow clouds, the fast
clouds can have an effective thickness of $\sim$ 2.5 kpc. Therefore, fast 
clouds can be part of the halo of the Galaxy. Alternative evidence 
supports the existence of atomic gas in the Galactic halo.
Albert \nocite{al83}(1983) selected lines of sight wherein a halo star 
and a nearby star
are aligned one behind the other. Absorption lines of TiII, CaII and 
NaI were measured towards these stars.
Though with a limited sample size of nine directions, the results of 
this study clearly indicate that the higher velocity absorption lines are 
seen only towards the distant star. She concludes that the high velocity 
tail seen in the optical line studies arise from the gas in the Galactic halo. 
Later, similar
studies (Danly, \nocite{danly89}1989; Danly et al, 
\nocite{danly92}1992;
Albert et al, \nocite{al94}1994; Kennedy et al, \nocite{gc01}1996,
\nocite{gc02}1998a, \nocite{gc03}1998b) confirmed this
result. The halo gas shows a larger spread in velocity, as compared to 
the gas
in the Galactic disk. Furthermore,
recent HI emission studies using the Green Bank
telescope have led to the discovery of a population of discrete HI 
clouds
in the Galactic halo with a velocity dispersion similar to that of the 
fast clouds
reported here (Lockman \nocite{jay02}2002).
The mean HI column density of these clouds was
estimated to be a few times 10$^{19}$ cm$^{-2}$.
He concludes that a cloud population with a line of
sight velocity dispersions of $\sigma_v$ $\sim$15 -- 20 km s$^{-1}$ is
capable of
explaining the observed velocity spread of these features. Lockman
\nocite{jay02}(2002) finds that many clouds in the halo have narrow 
line
widths implying temperatures below 1000K. In
addition, indications for a core-halo structure for these clouds was 
also
found. The fast clouds detected in the present survey are likely to
be part of the same halo gas detected in HI emission from Green Bank.
The fast clouds, once detected only in optical absorption lines,
have now been detected in both HI absorption and emission leading to a 
clearer picture of the interstellar medium.\\

\newpage

\noindent{{\textbf{Acknowledgements:}}} \\
\noindent{We wish to thank Shiv Sethi for useful
discussions related to the numerical techniques. We thank the referee,
 Miller Goss, for detailed comments and constructive criticisms
resulting in an improved version of this paper. We thank the staff
of the GMRT who made these observations possible. The GMRT is operated
by the National Centre for Radio Astrophysics of the Tata Institute of
Fundamental Research. This research has made use of NASA's Astrophysics
Data System.}


\label{lastpage}

\end{document}